\documentclass[12pt]{article}
\usepackage{amsmath}
\usepackage{graphicx}
\usepackage{geometry}
\usepackage{natbib}

\ifx\pdfoutput\@undefined\usepackage[usenames,dvips]{color}
\else\usepackage[usenames,dvipsnames]{color}
\IfFileExists{pdfcolmk.sty}{\usepackage{pdfcolmk}}{} 
\fi
\usepackage[plainpages=false,pdfpagelabels,pagebackref=false,naturalnames=true,hyperindex=true,pdftitle={The Implications of Interactions for Science and Philosophy},pdfauthor={Carlos Gershenson}]{hyperref}
\hypersetup{colorlinks=true,
urlcolor=Cerulean,linkcolor=BrickRed,citecolor=RoyalBlue,a4paper,
  pdfpagemode=None,
  pdfstartview=FitH}
\usepackage[all]{hypcap}

\usepackage[tight]{subfigure}

\begin{document}

\title{The Implications of Interactions\\ for Science and Philosophy}
\author{Carlos Gershenson\\
Computer Sciences Department,\\
Instituto de Investigaciones en Matem\'aticas Aplicadas y en Sistemas \\
Universidad Nacional Aut\'onoma de M\'exico\\
A.P. 20-726, 01000 M\'exico D.F. M\'exico\\
\href{mailto:cgg@unam.mx}{cgg@unam.mx} \
\url{http://turing.iimas.unam.mx/~cgg} 
}
\maketitle

\begin{abstract}
Reductionism has dominated science and philosophy for centuries. Complexity has recently shown that interactions---which reductionism neglects---are relevant for understanding phenomena. When interactions are considered, reductionism becomes limited in several aspects. In this paper, I argue that interactions imply non-reductionism, non-materialism, non-predictability, non-Platonism, and non-nihilism. As alternatives to each of these, holism, informism, adaptation, contextuality, and meaningfulness are put forward, respectively. A worldview that includes interactions not only describes better our world, but can help to solve many open scientific, philosophical, and social problems caused by implications of reductionism.
\end{abstract}

\section{Introduction}

Since the times of Galileo, Newton, Descartes, and Laplace, science has been reductionist \citep{Kauffman2008}. This implies that phenomena are studied by separating and simplifying their components to predict the future states of systems. An example of this classical scientific worldview, which still dominates science and philosophy, can be seen in textbook exercises. Systems are usually isolated, i.e. closed, ideal conditions are assumed, and elements are ``well behaved".  

As a scientific method, reductionism has been quite successful. It has lead to the development of electronics, the cure for most infectious diseases (thus dramatically increasing life expectancy worldwide), extraterrestrial exploration, controlled nuclear fission, mass production, etc. However, a long list of success stories does not imply that reductionism has not any limit. In this paper, it will be argued that recent advancements in science have made a reductionist worldview obsolete. This also implies changes in philosophy, which is based on the assumptions of classical reductionist science \citep{GershensonEtAl-PnC,HeylighenEtAl2007}. 

One of the main assumptions of reductionist science and philosophy is that the universe is predictable \citep{Gershenson:2011}, at least in theory. 
As it will be shown below, the predictability of the world is limited due to complexity.
Nevertheless, the predictability assumption has led science, spearheaded by physics, to take the implicit mission to find the ``true" laws ruling the universe. Under this assumption lies the ontological belief that there is one ``right" description of the universe, meaning that all other descriptions are wrong. This can be traced back to Plato. 

Ontological statements, i.e. about things as they ``really" are, independent of an observer, are fictitious. Once a statement---ontological or of any kind---is made, this has been made by an observer, turning it into epistemology, i.e. things as they are perceived and described. Even when reductionist philosophy has made ontological statements, e.g. the world is made of matter and energy, these have to be treated as epistemology. This opens the door for different descriptions of the world to co-exist without contradictions: they will be more useful for different contexts.
Reductionist epistemology is limited because the laws of physics cannot be related to the laws of biology. Even within physics, the Navier-Stokes equations of fluid motion cannot be reduced to quantum mechanics \citep[p. 17]{Kauffman2008}.
An encompassing epistemology will be able to relate different fields and context. A worldview based on information has this potential, since all epistemology can be treated as information \citep{Gershenson:2007}.

In the next section, an introduction to complexity and complex systems is given. The main section of the paper exposes the implications of interactions for science and philosophy: Non-reductionism, non-materialism, non-predictability, non-Platonism, and non-nihilism. Conclusions close the paper.

\section{Complexity}

The term complexity has been used in different contexts, so it can acquire a variety of meanings. Etymologically, complexity comes from the Latin \emph{plexus}, which means interwoven. Thus, a complex system can be understood as one in which the elements are difficult to separate. This difficulty arises from the \emph{interactions} between elements \citep{GershensonHeylighen2005}. Without interactions, elements can be separated. But when interactions are relevant, elements co-determine their future states. Thus, the future state of an element cannot be determined in isolation, as it co-depends on the states of other elements, precisely of those interacting with it.

We can find examples of complex systems all around us \citep{Bar-Yam1997,Mitchell:2009}: cells are composed of interacting molecules; brains are composed of interacting neurons, societies are composed of interacting individuals, ecosystems are composed of interacting species.

\begin{figure}
     \centering
     \subfigure[]{
          \label{fig:GoLA}
          \includegraphics[width=.45\textwidth]{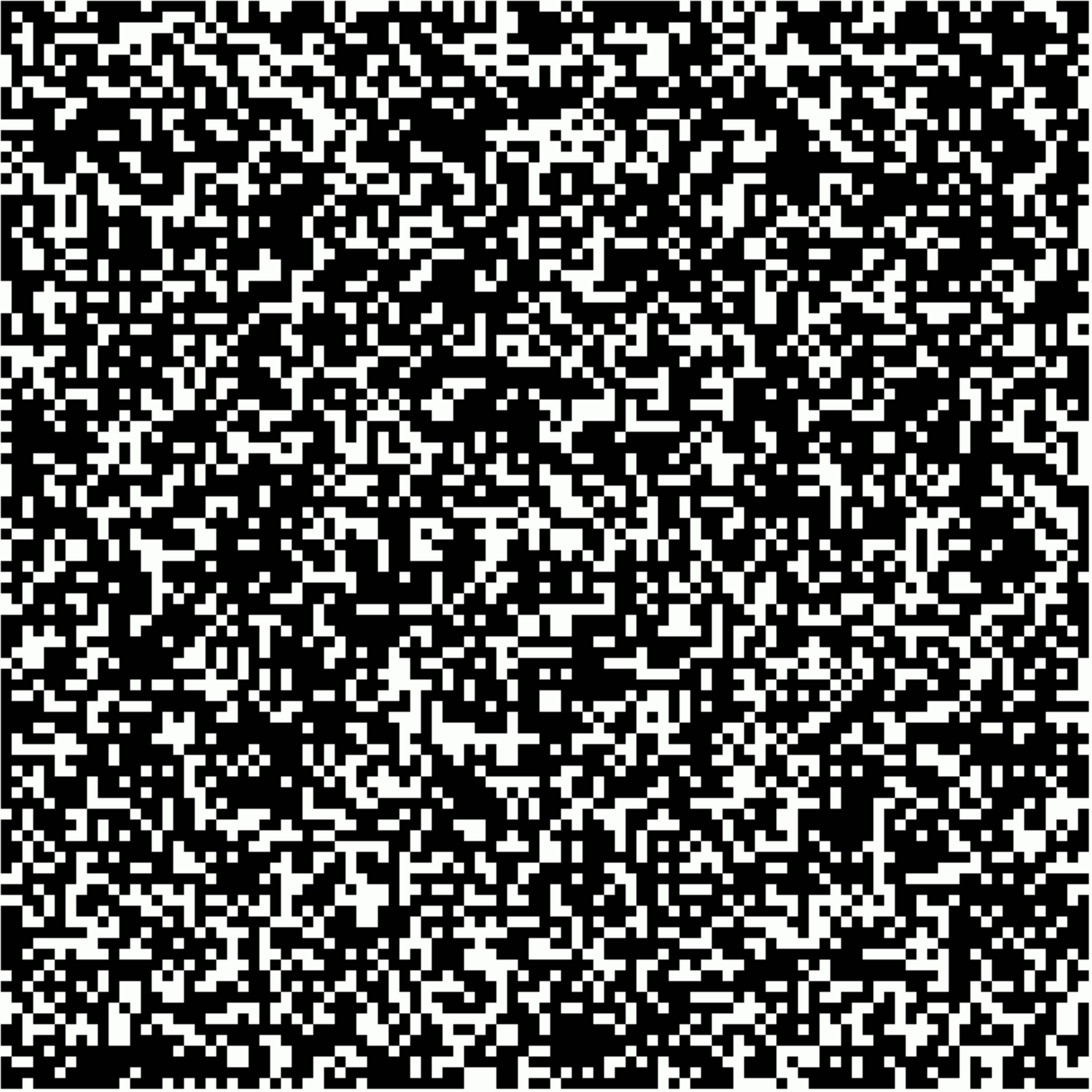}}
     \subfigure[]{
          \label{fig:GoLB}
          \includegraphics[width=.45\textwidth]{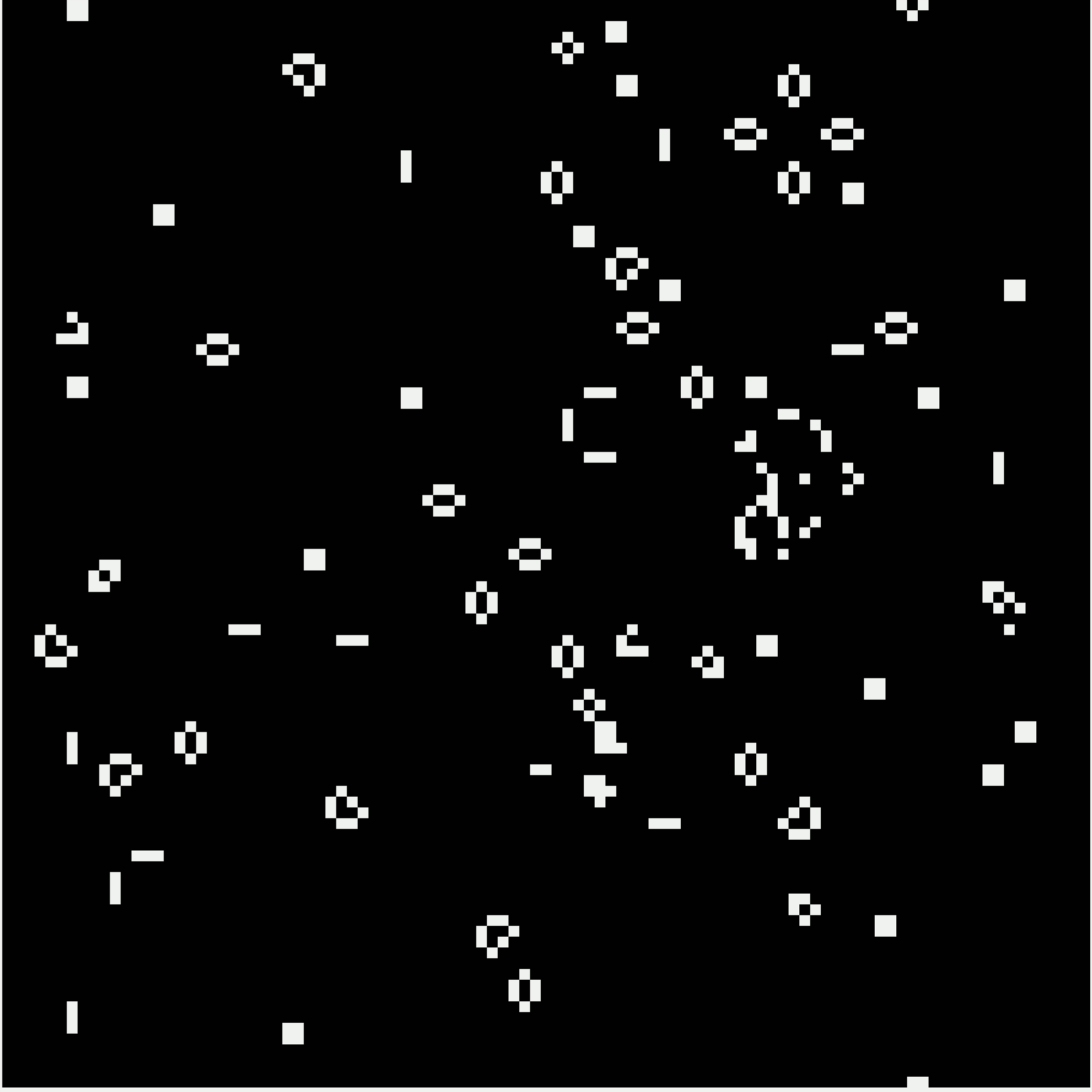}}
\\
     \subfigure[]{
          \label{fig:GoLC}
          \includegraphics[width=.45\textwidth]{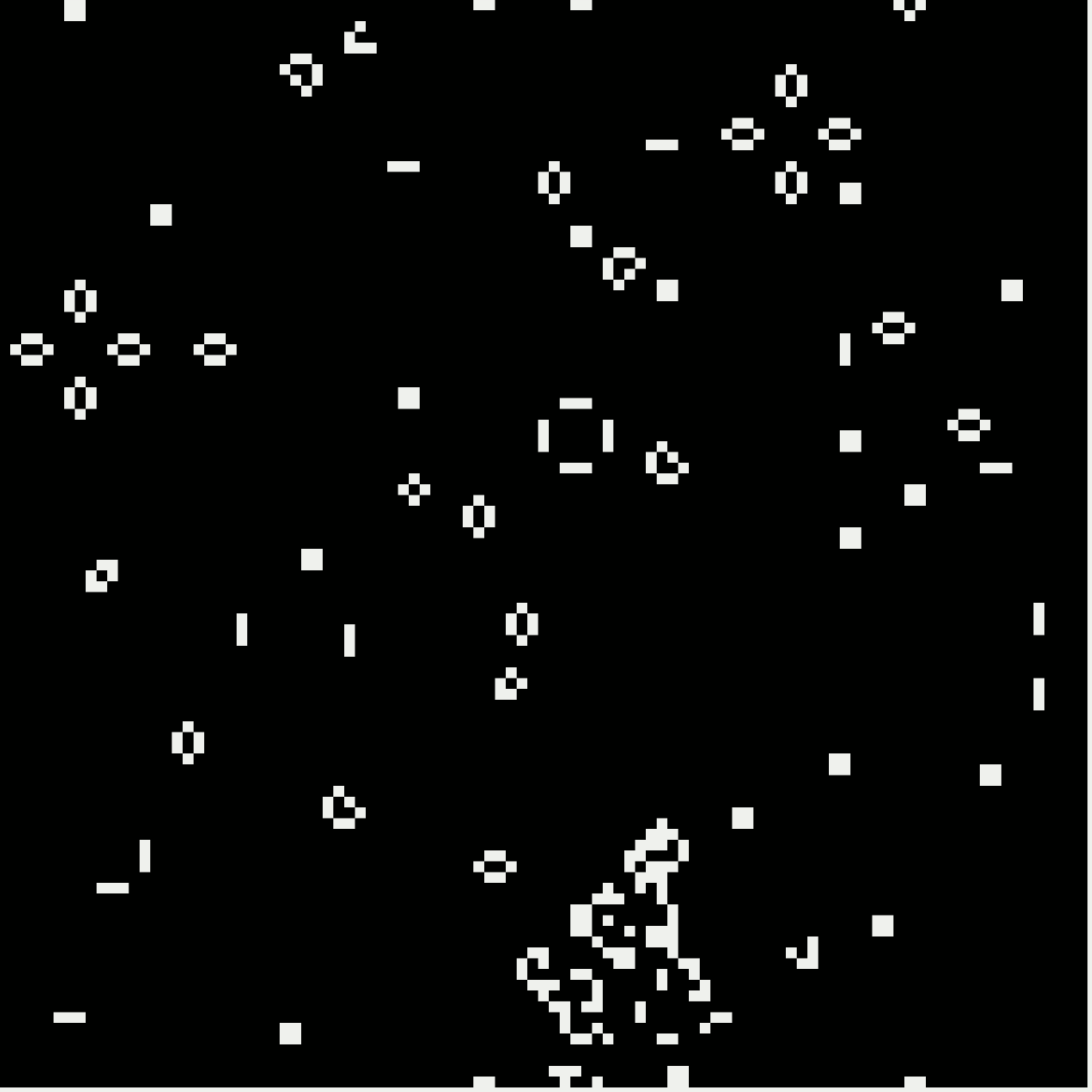}}
     \subfigure[]{
          \label{fig:GoLD}
          \includegraphics[width=.45\textwidth]{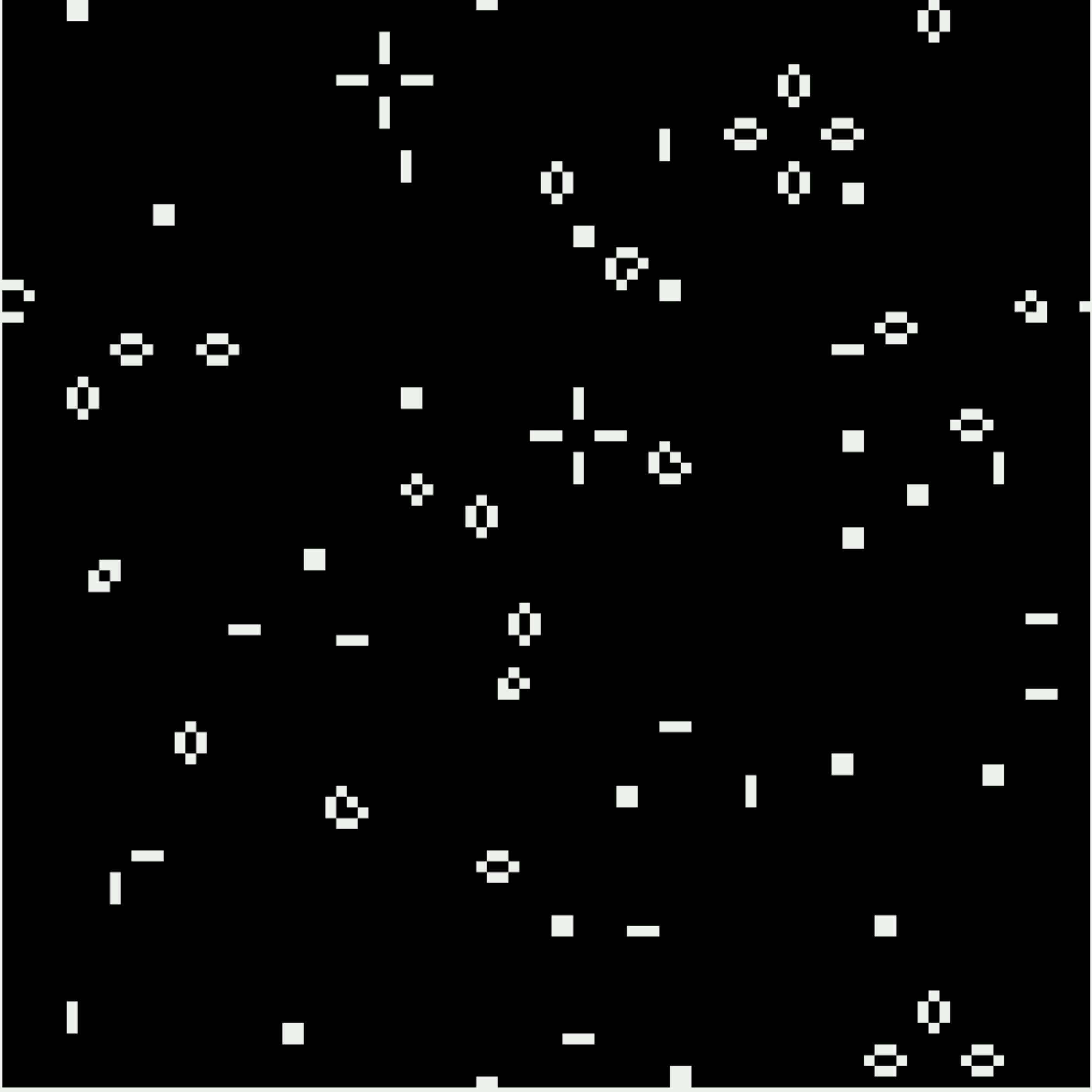}}

     \caption{ Evolution of GoL starting from a random initial configuration (A), where white cells are ``alive" and black cells are ``dead". After 410 steps (B), some stable structures have formed, but there are still some zones with activity. After 861 steps (C), some of the stable structures have been destroyed by the dynamics and some new ones have been created. There is still some activity on the bottom of the grid. After 1416 steps (D), the dynamics have stabilized, with some static and some oscillatory structures. Images created with NetLogo \citep{Wilensky1999}. }
     \label{fig:GoL}
\end{figure}

A classical example can be seen with Conway's ``Game of Life" (GoL) \citep{BerlekampEtAl82}. This consists of a two-dimensional Boolean cellular automaton \citep{vonNeumann1966,Wolfram1986}, i.e. a grid where different cells can take two values: ``dead" (0) or ``alive" (1). The future state of cells depends on how many of their eight neighbors are alive. If a cell is alive, it will remain alive ($1\rightarrow1$) if it has two or three neighbors, but die ($1\rightarrow0$) if it has less than two or more than three. A dead cell will become alive ($0\rightarrow1$)  if there are exactly three living neighbors around it, and otherwise will remain dead ($0\rightarrow0$). These are very simple rules, but lead to very complex dynamics. Figure \ref{fig:GoL} shows an example run from random initial conditions.

GoL contains an immense zoo of structures: stable patterns, oscillatory patterns, ``gliders" (structures with diagonal locomotion), ``spaceships" (structures with longitudinal locomotion), eaters (structures that destroy other structures), breeders (structures that generate other structures), etc\footnote{To explore GoL, the reader can download the Golly application at \url{http://golly.sourceforge.net/}}.
The dynamics of GoL are so rich that a Turing machine---a theoretical construct that defines universal computation---can be built within it \citep{BerlekampEtAl82}. In other words, the simple rules of GoL can produce dynamics rich enough to compute all (theoretically) computable functions. How can this be? If we attempt to describe all the richness of GoL examining different cells in isolation, not much will come out of it. If the ``laws" of GoL are so simple, is it possible to predict all the richness of its dynamics? No, because it is computationally irreducible \citep[pp. 737-750]{Wolfram:2002}. In other words, the shortest path to calculate a future state of a system is by computing it. Why does analysis fail? Because of \emph{interactions}. Interactions generate \emph{novel} information that is not present in initial and boundary conditions. Even if we know the laws of a deterministic system and an initial state, we cannot predict a future state before we ``run" the system. Interactions lead to complexity, and complexity leads to irreducibility. Thus, one might ask, how can we decide whether a system is complex or not?

Whether we describe a living cell as complex (billions of heterogeneous, interacting molecules) or not (a Boolean variable: 0 if dead, 1 if alive) is not an ontological decision, but an epistemological one. Both descriptions are possible, the preferable one will depend on the context in which it is used. Thus, all systems are potentially complex or simple, depending on how we decide to describe them \citep{GershensonHeylighen2003a,GershensonHeylighen2005}.

Nevertheless, interactions are as real as matter and energy. The fact that most interactions cannot be seen or touched or smelled does not mean that they are not there. This does not suggest that interactions have to be physical, but the opposite: matter and energy are epistemic \citep{Gershenson:2007}. 
The non-reducibility of interactions to the language of physics---where matter and energy are considered as basic elements---is a strong reason to speak in terms of information, since the language of physics does not translate well to all scales, while information does \citep{vonBaeyer2004}.

\section{Implications of Interactions}

Since reductionism neglects interactions, and these are relevant for the description of complex systems, there are several implications of interactions for science and philosophy. Again, these are epistemic, not ontological. As discussed above, every scientific and philosophical discourse is epistemic. Even ontology is epistemic, as it is described by an observer. 

In the next subsections, holism is defended as an alternative to reductionism, informism as an alternative to materialism, adaptation as an alternative to prediction, contextuality as an alternative to Platonism, and meaningfulness as an alternative to nihilism. 

\subsection{Non-reductionism}

Reductionism in physics assumes that ``reality"---whatever it might be---lies only in the realm of physics, and that all the phenomena we observe in chemistry, biology, sociology, psychology, philosophy, etc. are ``nothing more" than the aggregation of whatever ``elementary" particles are in vogue\footnote{In ancient Greece, atoms were defined as the basic elements of matter, literally meaning ``indivisible". After atoms were split into protons, electrons, and neutrons, subatomic particles were considered to be elementary. Nowadays, the standard model of physics assumes that elementary particles are fermions (quarks, leptons, and bosons) and antifermions. Theoretically, nothing prevents these elementary particles from being divided. Experimentally, however, division of fermions would require very high energies, which are not yet available.} \citep{Weinberg:1994}. Few people defend this view strongly, but most of science and philosophy falls within the assumptions of reductionism.

This reductionist view can be criticized on several fronts. For example, the standard model is the most widely accepted physical theory, but not the only one, and there are still experimental holes waiting to be filled. We are far from having a complete view about the reality that reductionism promotes. 
Still, a reductionist might argue that this is just a matter of time, even if the standard model has to be modified, or a better theory is developed, soon we will be able to understand all of the universe, parting from a grand unified theory (GUT), as the standard model attempts to become.

However, interactions show that a GUT is a hopeless task, in theory and practice. It might explain much of physics, but never all phenomena that we perceive. This is because interactions generate novel information, so even with the complete list of all the ``laws" of the physical universe (which is not seen on the horizon), it will not be possible to deduce from these life, cognition, or meaning. This is simply because the information generated by interactions is neither in the laws nor in initial conditions: it is generated ``on the run". The universe is computationally irreducible \citep[pp. 737-750]{Wolfram:2002}.

The novel information generated by interactions can be described as \emph{emergent} \citep{Emergence:2007}. This implies that novel properties arise from the interactions between components of a system, i.e. emergent properties are not present in the components and cannot be reduced to them. This is precisely because of interactions, since these are not predefined in the behavior of the elements. The response to an interaction can be defined, as in the Game of Life, but one does not know the precise order in which interactions will take place, which can alter the future state of the system. Emergent properties are pervasive, whether one wants to accept them or not: a cell is composed of molecules, but life cannot be reduced to chemistry; a brain is composed of neurons, but cognition cannot be reduced to neurophisiology. An ecosystem is composed of species, but ecology cannot be reduced to ethology. Reductionism has promised but not delivered.

This rejection of reductionism naturally leads to \emph{holism} (Aristotelian, not new age), which can be summarized by the phrase ``the whole is greater than the sum of its parts". This implies that the descriptions of phenomena from different fields cannot be reduced to physics. In other words, physics has lost its hegemony in science, putting it on par with all other disciplines. There is no ``ruling" discipline, as each discipline studies different phenomena using different languages in different contexts.

One of the main obstacles of reductionism is the inability to reduce meaning to physics \citep{Neuman:2008}. This is because the realm of physics is suitable to describe matter and energy, but not meaning, which finds itself at home in the realm of information.

\subsection{Non-materialism}

Since reductionism assumes that reality is within physics, and that the components of the physical universe are matter and energy, many have been led to believe that the only real ``stuff" is matter. Materialism implies that ideas and emotions are not real. This led many philosophers---including Descartes---to propose dualism as an alternative to describe mental phenomena. However, having two or more ontological categories has the problem of relating phenomena belonging to different categories.

Since ontological statements are finally epistemic, matter and energy should be considered as another description of the universe. Interactions as real as matter and energy, as these are just different descriptions of perceived phenomena. Matter and energy cannot be used to describe all perceived phenomena, but \emph{information} can be used \citep{Gershenson:2007}. It is natural to describe interactions as information, but matter and energy can also be seen as particular types of information. The same holds for life, cognition, and other phenomena that cannot be reduced to matter and energy.

The rejection of materialism as an appropriate ontology/epistemology invites us to explore \emph{informism} as an alternative to describe our world. Thus, instead of attempting to reduce information and meaning to matter and energy, we can express matter and energy as one type of information. Informism is a monism where phenomena at all scales can be related using the same language and a single ontological/epistemological category  \citep{Gershenson:2007}.

Economy offers a clear example of the limits of materialism. Agricultural and industrial economies relied in the exchange of mostly material and energetic goods. However,  information economies \citep{Toffler:1980} exchange non-material goods, such as software, services, and knowledge in general. Not only we give monetary value to non-material goods, but the concept of money itself is non-material. What is the value of a dollar? of a gold coin? of a cacao seed? of a Google share? of a song? of an advice? Economic value is not determined by matter, e.g. coins of devaluated currencies might be more valuable by weight than by currency. Or vice versa: a coin can be worth more by currency than by weight. Economic value is \emph{agreed} upon by the \emph{interactions} between different stakeholders. A materialist economy was useful in the Industrial Age, where values did not change considerably over long periods of time. However, materialism in economy has been obsolete for a long time, as values of non-material goods can change drastically within minutes. It is clear that money is non-material. Would anyone argue that money is not real, while having causal effects on our world at all scales?

\subsection{Non-predictability}

As mentioned in the Introduction, one of the main assumptions of classical science has been that the world is predictable. Since the end of the nineteenth century, the predictability of the world was questioned \citep{Morin2006}, starting with the three-body problem studied by Poincar\'e. Nevertheless, many scientists and philosophers still assume that the world is and should be predictable. 
Since the uncertainty principle of quantum mechanics is believed to hold only at microscopic scales, the practical effects of non-determinism for the predictability of the world were neglected. 
It was until the second half of the twentieth century that deterministic chaos gave an initial blow to the predictability assumption \citep{Gleick:1987}. Deterministic chaos occurs when there is an exponential divergence of similar trajectories. In other words, very similar states will quickly tend to very different states. This is also called sensitivity to initial conditions. Even when the ``laws" or equations describing a phenomenon are known, there is never enough precision to make long term predictions, as it is the case with the weather. 

However, some people believe that this limit to predictability is only practical, as in theory chaotic phenomena are predictable when an infinite precision is used.

Nevertheless, complexity has shown that predictability is limited. Since novel information is generated by interactions, all information cannot be known beforehand (from rules and initial conditions). Thus, the future state of a system cannot be reduced, it has to be ``run" before one can expect an answer, i.e. there is computational irreducibility. Certainly, this has led people to build models that can run in computers faster than phenomena. However, this predictability is as limited as the model is simplifying. And the less simplifying a model is, the longer it will take to run.

Actually, the non-predictability of the weather depends more on interactions than on deterministic chaos, as perturbations from ``outside" the modeled area can change considerably the trajectories of the troposphere. It should be mentioned that chaos does not imply complexity, nor vice versa. There can be non-chaotic complex systems (e.g. some elementary cellular automata (ECA) \citep{WuenscheLesser1992}), chaotic simple systems (i.e. without interactions, e.g. logistic map), chaotic complex systems (e.g. other ECA), and non-chaotic simple systems (the velocity of an object in free fall in a planet without atmosphere, before landing).

I am not suggesting to abandon all hope of prediction, I am suggesting to abandon the assumption that we can and should predict every phenomena. In many cases, predictability can be partial, depending on the scale at which prediction is made. Again, the weather is a good example, since it is partially predictable in a short time scale, which is better than nothing. Nevertheless, if we are aware that prediction is limited, then we can complement it with \emph{adaptation} \citep{GershensonDCSOS}. In this way, if we build a system that has to deal with a complex environment---which implies limited predictability due to novel information generated by interactions---when changes occur in the environment (and its interactions with the system), the system will be able to adapt, i.e. modify its behavior or functioning, without the need of human intervention. Living systems constantly do this, so different adaptation mechanisms can be inspired by them.

\subsection{Non-Platonism}

Plato presents the allegory of the cave in The Republic, book VII. Briefly, the allegory tells of people living chained inside a cave, where they can only see shadows of real objects. However, there \emph{is} a true reality outside the cave, which can be perceived by people who have been freed. This allegory illustrates the desire of science and philosophy to find the ``true" laws and descriptions of the world. 

As Wittgenstein noted, all descriptions of the world have to go through a language. Since there is no universal language, different descriptions of the same phenomenon will be possible. Which description is the ``true" or ``real" one? The inability to decide this question led some (but by no means all) postmodernists to promote an ``anything goes" approach, where all descriptions are as valid as any other. However, experience tells us otherwise: some are right, some are wrong. Again, how can we decide which is which?

A solution can be illustrated with the ``silly theorem problem": For any silly theorem (e.g. 1+1=10), one can find an infinitude of sets of axioms under which the silly theorem will be consistent (e.g. in binary, 1+1=10, since two is represented as one-zero). How come we don't have silly theorems in mathematics? Because \emph{experience} helps us in deciding which ones are \emph{useful} and which are not \citep{Chaitin:2004}. Certainly, many mathematicians do not like being told that mathematics is experience-based, as it adds a whiff of despised subjectivity to their field.

However, just as we are not dealing with total objectivity, we are not dealing with total subjectivity. A more precise term would be \emph{contextuality} \citep{Gershenson2002ua}. This implies that all statements and descriptions are made within a context. Within that context, and only within that context, one can speak about the ``truth" of a statement. When the context changes, rules might change. If a context is agreed upon, then a discussion can take place \emph{within} that context.

If we wanted to have a \emph{complete} description of a phenomenon, we would require to have a complete description of all of its potential contexts. Since the number of all potential contexts is infinite, a complete description of a phenomenon is not possible. Thus, all descriptions of phenomena are incomplete and contextual. 

Having only contextual descriptions, the search for absolute truths (in all possible contexts) is futile. People who believe that there is an absolute truth---and/or that the mission of science is to find this truth---are Platonist, whether they are aware of it or not. Platonism is not \emph{wrong}, but is limited to a specific context that rejects different descriptions of phenomena. Contextuality is not \emph{right}, it just allows for different descriptions to coexist. As Heraclitus of Ephesus noted, everything changes. Thus, different descriptions of the world must change, as their contexts also change. 

As an example, imagine a ball that is half white an half black. Different people see it from a different perspective. Of which color is the sphere, perceived as a circle? It will depend on the perspective of the observer. Some will say black, some will say white, some will say half and half, or 80\% and 20\%. One cannot average answers, as a majority of observers could have a biased perspective. If we can change perspective---which Platonism rejects (once outside the cave)---more facets of the sphere could be perceived, by going from two to three dimensions. A similar thing occurs with any perceived phenomena, only that there is an infinite number of dimensions (contexts) which can give novel information about the phenomena. We will never reach the absolute perception of any phenomenon, as there will be always new dimensions/contexts to be considered. 

\begin{figure}[htbp]
\begin{center}
 \includegraphics[width=.8\textwidth]{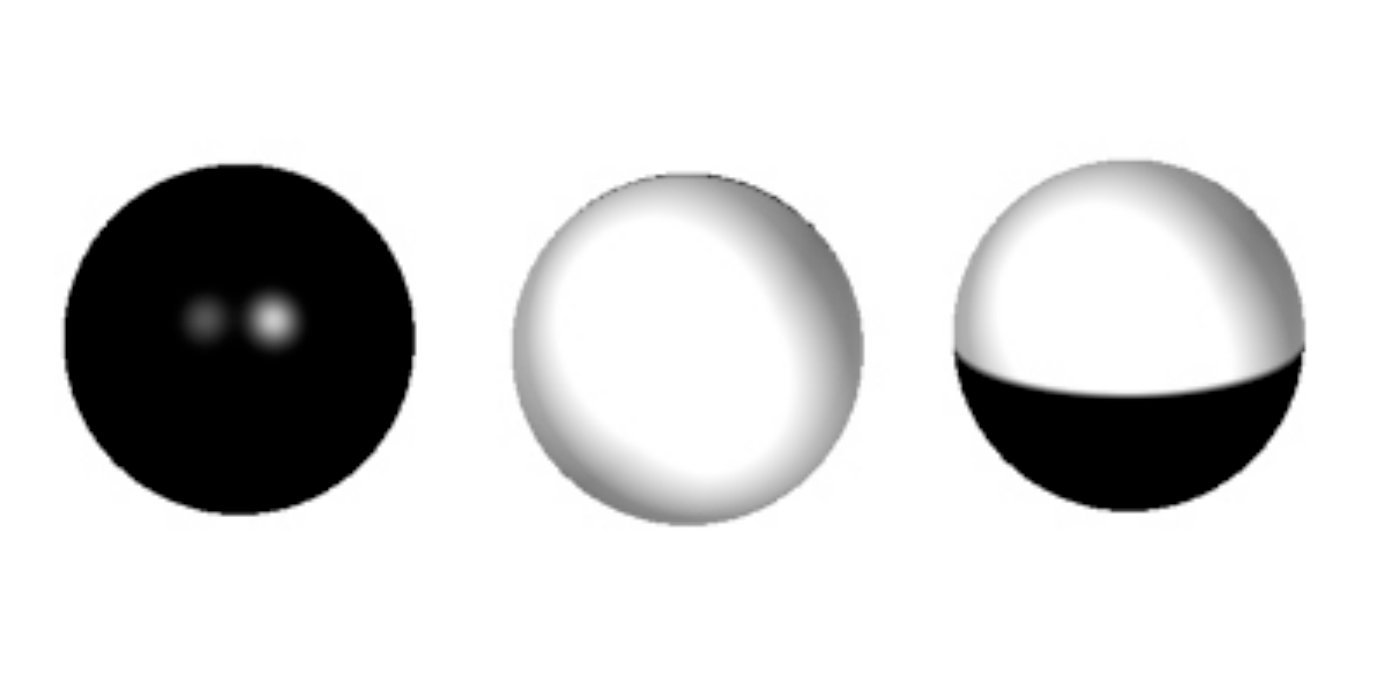}
\caption{Of which color is the circle? It depends on the perspective from which it is observed.}
\label{fig:balls}
\end{center}
\end{figure}

\subsection{Non-nihilism}

Reductionism assumes that all of the universe can be explained with a grand unified theory (GUT), yet to be found, but rooted in physics. This scientific idea has strong implications in philosophy. If we are interested in discussing the meaning of life, free will, ethics, aesthetics, and many other features of the human condition, then reductionism tells us that there is nothing special about humanity, since humanity can be reduced---somehow---to the laws of physics. This negation of the meaning of human life is the basis of nihilism. 

However, since there is novel information and meaning in interactions, there is always room for amazement. The story is not told or predefined, since new meanings can arise. Even the same phenomena can acquire new meanings, as they are perceived in novel contexts, i.e. new interactions  \citep{Gershenson:2007}. There is not only \emph{a} meaning in human life. There are potentially \emph{infinite} meanings, waiting to be discovered.

This \emph{meaningfulness} offers a scientifically-based spirituality, which traditional science often rejects. Since rejecting spirituality does not replace it, meaningfulness brought by interactions can fill the void left. 

\section{Conclusions}

Reductionism in science studies phenomena in isolation for convenience. However, nothing is isolated. Everything is related by a web of interactions. Whether we might decide to neglect interactions does not make them to go away. We might to use an isolating context for convenience in the understanding of a particular phenomenon. But the ``big picture" will not be seen within that context. Once we move to a context where we describe all phenomena as interacting and related, a broader understanding of our world will lead to less conflicts. This is because when an interacting world is perceived, the separation between different phenomena is understood as artificial. If we see ourselves as interrelated, as part of the same whole, we will understand that positive actions and interactions will lead to positive consequences, while negative actions and interactions will lead to negative consequences. This encompassing view is consistent with Buddhist philosophy \citep{Nydahl:2008}. 

Interactions and complexity demand changes in our worldviews \citep{Aerts:1994,Vidal:2008} to fit better with our contexts.
In this paper, several arguments against reductionism were given. Even when the alternatives presented here are more appropriate for our current contexts, these will not be infallible. There will never be a final theory. Ideas and descriptions will keep on evolving, as our contexts do. It is consequence of our interacting world.

\section*{Acknowledgments}

I should like to thank all the people who have contributed over the years to the development of the ideas presented here. This work was partially supported by SNI membership 47907 of CONACyT, Mexico. 

\bibliographystyle{cgg}
\bibliography{carlos,complex,information,evolution,philosophy,RBN,sos,traffic}

\end{document}